\documentclass[aps,prl,twocolumn,groupedaddress,showpacs]{revtex4}
\usepackage{graphicx}

\newcommand{\delete}[1]{}

\newcommand{\be}{\begin{equation}}
\newcommand{\ee}{\end{equation}}

\def\beq{\begin{equation}}
\def\eeq{\end{equation}}
\def\bea{\begin{eqnarray}}
\def\eea{\end{eqnarray}}
\def\ba{\begin{array}}
\def\ea{\end{array}}

\begin{document}

\title{Resonant detection of axion mediated forces with Nuclear Magnetic Resonance}
\author{Asimina Arvanitaki$^1$, Andrew A. Geraci$^2$}
\affiliation{$^1$Perimeter Institute for Theoretical Physics, Waterloo, ON N2L 2Y5, Canada}
\affiliation{$^2$Department of Physics, University of Nevada, Reno, NV 89557, USA}

\date{\today}
\begin{abstract}
We describe a method based on precision magnetometry that can extend the search for axion-mediated spin-dependent 
forces by several orders of magnitude.  By combining techniques used in nuclear magnetic resonance and short-distance tests of gravity, our approach can substantially improve upon current experimental limits set by astrophysics, and probe deep into the theoretically interesting regime for the Peccei-Quinn (PQ) axion. Our method is sensitive to PQ axion decay constants between $10^9$ and $10^{12}$ GeV or axion masses between $10^{-6}$ and $10^{-3}$ eV, independent of the cosmic axion abundance. 
\end{abstract}


\maketitle

{\it{Introduction.}} Axions are CP-odd scalar particles that are present in a variety of theories beyond the Standard Model. Their mass is protected by shift symmetries so they remain naturally light and their couplings to matter are very suppressed. In string theory in particular, they naturally arise in compactifications with non-trivial topology \cite{Svrcek:2006yi, axiverse}. The mass of string axions exponentially depends on the parameters of the theory, and can be as small as the Hubble scale. The most famous axion is the Peccei-Quinn (PQ) axion \cite{axion} whose presence explains the smallness of the neutron's electric dipole moment and has been the main focus of experimental searches since it was proposed over 30 years ago. Its mass is generated by non-perturbative QCD effects. If lighter than $10^{-5}$~eV, the PQ axion becomes an excellent dark matter candidate. In laboratory experiments, axions can generate novel spin-dependent short-range forces between matter objects \cite{Moody:1984ba}.

In this paper, we propose a magnetometry experiment based on nuclear magnetic resonance (NMR) that searches for axion mediated CP-violating forces with a range between $\sim 100~\mu m$ and  $\sim 10~cm$ or axion masses between $\sim10^{-6}$~eV and $\sim10^{-3}$~eV. Our proposal is based on the resonant coupling between the rotational frequency of a source mass and an NMR sample 
with a matching spin precession frequency. Similar techniques involving resonant excitation are used in short-distance gravity experiments \cite{stanford, adelberger, price}. In the presence of an anomalous CP-violating interaction between the source mass and the NMR detector, the spins in the NMR material resonantly precess off the axis of polarization. This change in the magnetization can be read by a superconducting quantum interference device (SQUID).

There are already several methods based on precision magnetometry to look for such spin-dependent short range forces, see for example \cite{Vasilakis:2008yn,gsgpgermany,snow} (for a summary of recent results see Ref. \cite{Raffelt:2012sp}).  In previous experiments, shifts of the spin-precession frequency are observed as matter objects are brought into and out of proximity with a sample. Our setup is different from previous approaches as the detection technique is based on a resonant effect, where the source mass {\it{itself}} is moved periodically at the Larmor frequency in order to drive spin precession in the NMR medium. This helps reduce several systematics and at the same time takes advantage of the enhancement of the signal due to the high spin density of the NMR material ($\sim 10^{21}~cm^{-3}$) and the quality factor of the NMR sample which can be as high as $10^6$.

In the following, we show how the proposed setup can probe both the monopole-dipole and the dipole-dipole coupling of axions at a level that is competitive with astrophysical bounds. The experiment can eventually be up to 8 orders of magnitude more sensitive than current approaches and can bridge the gap between astrophysical bounds and cosmic PQ axion searches \cite{ADMX,casper}, without requiring that the axion is dark matter or the need to precisely scan over its mass.

{\it{Axion-mediated forces.}}   The interaction energy between particles due to monopole-dipole axion exchange as a function of the distance $r$ is:
 \be
 U_{sp}(r)=\frac{\hbar^2 g_s g_p}{8 \pi m_f}\left( \frac{1}{r \lambda_a}+\frac{1}{r^2}\right) e^{-\frac{r}{\lambda_a}} \left(\hat \sigma \cdot \hat r \right),
 \ee
where $m_f$ is the fermion mass, or in the case of dipole-dipole axion exchange:
 \begin{eqnarray}
 U_{pp}(r)=\frac{\hbar^3 c}{16 \pi}\frac{ g_{p_1} g_{p_2}}{m_{f_1} m_{f_2}} (\left(\hat \sigma_1 \cdot \hat \sigma_2\right)\left( \frac{1}{r^2 \lambda_a}+\frac{1}{r^3}\right) e^{-\frac{r}{\lambda_a}}\\ \nonumber
- \left(\hat \sigma_1 \cdot \hat r \right)\left(\hat \sigma_2 \cdot \hat r\right) \left( \frac{1}{r \lambda_a^2}+\frac{3}{r^2 \lambda_a}+\frac{3}{r^3}\right)  e^{-\frac{r}{\lambda_a}})
 \end{eqnarray}
 The range of interaction is set by the mass of the axion $\lambda_a=\frac{\hbar c}{m_a }$. It is convenient to write interactions that involve spins ({\it i.e}. dipoles) using the axion potential
 \be
 \label{eq:axionpotential}
-\vec \nabla V_a(r)\cdot \hat \sigma_2,
 \ee
 where $V_{a_s}(r) = \frac{\hbar^2 g_s g_p}{8 \pi m_f}\frac{e^{-\frac{r}{\lambda_a}}}{r}$, for monopole-dipole interactions, or $
 V_{a_p}(r)=\frac{\hbar^3 c}{16 \pi}\frac{ g_{p_1} g_{p_2}}{m_{f_1} m_{f_2}}(\hat \sigma_1 \cdot \hat r)\left(\frac{1}{r^2} + \frac{1}{\lambda_a r}\right)e^{-\frac{r}{\lambda_a}}$, if an axion can be exchanged between two spins.
 For the PQ axion $g_s$ and $g_p$ are directly correlated to the axion mass as they are fixed by the axion decay constant $f_a$:
 \begin{eqnarray}
& 6\times 10^{-27} \left( \frac{10^9~GeV}{f_a}\right)\lesssim g_s \lesssim  10^{-21}  \left( \frac{10^9~GeV}{f_a}\right),\\
  &g_p= \frac{C_f m_f}{f_a}= C_f  10^{-9}   \left( \frac{m_f}{1~GeV}\right)\left( \frac{10^9~GeV}{f_a}\right),\mbox{~and}\\
 & m_a= 6 \times 10^{-3}\mbox{eV} \frac{10^9~\mbox{GeV}}{f_a}.
 \end{eqnarray}
The scalar coupling of the PQ axion is indirectly constrained by EDM searches and the lower bound is set by the amount of CP violation in the Standard Model \cite{scalaraxioncoupling}. There are large uncertainties in the QCD matrix elements involved in the calculations of this coupling and further study is required through lattice simulations. In the PQ axion coupling to spin, $C_f$ is a model dependent constant typically expected to be $ \mathcal{O}(1)$ \cite{PDG} and in what follows we assume $C_f=1$ for simplicity. The axion decay constant is constrained to be $10^9~\mbox{GeV} \lesssim f_a \lesssim 10^{17}~\mbox{GeV}$. Both these bounds on $f_a$ are set by astrophysics; the lower bound comes from red giant cooling and SN1987a, while the lesser known upper bound on $f_a$ arises because the wavelength of a large $f_a$ PQ axion is of order the size of stellar mass black holes. If such an axion existed it would have caused these black holes to spin down through the superradiance effect \cite{axiverse, BH}. They are thus excluded by the observation of several near extremal black holes.

Eq. \ref{eq:axionpotential} shows that the axion generated potential by an unpolarized or polarized mass acts on a nearby fermion just like an effective magnetic field of size and direction given by $\vec B_{\rm{eff}} = \frac{2 \vec \nabla V_a(r)}{\hbar \gamma_f}$,
where $\gamma_f$ is the fermion gyromagnetic ratio. This field is different from an ordinary EM field -- it couples to the spin of the particle, is independent of the fermion's magnetic moment, and different for nucleons and electrons. It also does not couple to ordinary angular momentum. Therefore, it crucially is not screened by magnetic shielding.


\begin{figure}[!t]
\begin{center}
\includegraphics[width=0.5 \columnwidth]{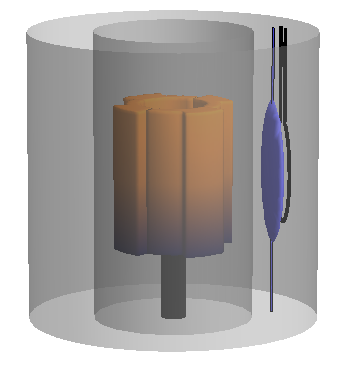}
\caption{ A source mass consisting of a segmented cylinder with $n$ sections is rotated around its axis of symmetry at a fixed frequency $\omega_{\rm{rot}}$, which results in a resonance between the frequency $\omega = n \omega_{\rm{rot}}$ at which the segments pass near the sample and the resonant frequency $2 \vec \mu_{N} \cdot \vec B_{ext} /\hbar $ of the NMR sample. The NMR sample has an oblate spheroidal geometry to minimize magnetic gradients while allowing close proximity to the mass. Superconducting cylinders screen the setup from the environment and the NMR sample from the source mass. }
\label{Fig:Experiment}
\end{center}
\end{figure}


{\it{Experimental Setup.}} Our proposed setup is schematically drawn in Fig. \ref{Fig:Experiment}. A quartz vessel containing hyperpolarized $^3$He gas is placed next to a segmented cylinder that acts as a source mass. The cylinder consists of either high density unpolarized material (e.g. tungsten) or material with a net electron or nuclear spin polarization, and is rotated around its axis of symmetry with a frequency $\omega_{\rm{rot}}$. To screen background electromagnetic fields, a superconducting niobium cylindrical shell is placed between the cylinder and $^3$He sample. The use of superconductors eliminates the magnetic field noise associated with Johnson noise near the surface of conducting materials \cite{varpula}. An axion with a Compton wavelength smaller than $R$ will generate a potential a distance $r$ from the surface of the cylinder given by $
V_{a_s} (r)=\hbar^2 \frac{g_s g_{p_N}}{2 m_N} \lambda_a^2 n_N e^{- \frac{r}{\lambda_a}}$,
if the axion has a monopole coupling to nucleons, where $m_N$ and $n_N$ are the nucleon mass and density of the material, respectively. If the axion has a dipole coupling to nucleons or to electrons and the polarization of the source mass is perpendicular to the axis of rotation, then $
V_{a_p} (r)=\hbar^3 c \frac{g_{p_f}g_{p_N}}{4 m_f m_N} \lambda_a n_s e^{-\frac{r}{\lambda_a}}$. Here $n_s$ is the polarized spin density in the material.

A spin polarized nucleus near this rotating segmented cylinder will feel an effective magnetic field $
B_{\rm{eff}}= \frac{1}{\hbar \gamma_N} \nabla V_a(r) (1+\cos(n \omega_{\rm{rot}} t)),$
where $\gamma_N$ is the nuclear gyromagnetic ratio and $n$ is the number of segments. Here we assume the NMR sample thickness is of order the axion Compton wavelength. This effective magnetic field is parallel to the radius of the cylinder. An NMR sample with net polarization parallel to the axis of the cylinder and a Larmor frequency $2 \vec \mu_N \cdot \vec B_{\rm{ext}} / \hbar = \omega$ determined by an axial field $B_{\rm{ext}}$ will develop a magnetization perpendicular to its polarization of magnitude:
\be
M(t)\approx \frac{\hbar}{2} n_s p \mu_N \gamma_N B_{\rm{eff}} t \cos(  \omega t),
\ee
where $p$ is the polarization fraction and $\mu_N$ is the nuclear magnetic moment. This polarization grows linearly with time until $t\sim T_2$ where $T_2$ is the transverse relaxation time of the sample. $M(t)$ can be detected by a SQUID magnetometer with its pickup coil axis oriented radially.

The main fundamental limitation comes from transverse projection noise in the sample itself $\sqrt{M_N^2} = \sqrt{\frac{\hbar \gamma n \mu_{^3\rm{He}}T_2 }{2V}}$ and the minimum transverse magnetic resonant field this setup is sensitive to is given by:
\begin{eqnarray}
\label{bmin}
&&B_{\rm{min}}\approx p^{-1}\sqrt{\frac{2\hbar}{n_s  \mu_{^3\rm{He}} \gamma V T_2}} =10^{-20} \frac{T}{\sqrt{Hz}} \times \\ \nonumber
&& \left(  \frac{1}{p} \right) \left( \frac{1~\mbox{cm}^3}{V}\right)^{1/2}\left( \frac{10^{21}~\mbox{cm}^{-3}}{n_s}\right)^{1/2}\left( \frac{1000~\mbox{sec}}{T_2}\right)^{1/2}.
\end{eqnarray}
Here $V$ is the sample volume, $\gamma$ is the gyromagnetic ratio for $^3$He $= (2\pi) \times 32.4$ MHz/T, and $\mu_{^3\rm{He}} = -2.12 \times \mu_n$ is the $^3$He nuclear moment \cite{3He}, where $\mu_n$ is the nuclear Bohr magneton. The equation above shows where the tremendous boost in sensitivity lies. First, the resonant enhancement of the signal gives rise to an increase of sensitivity because of an effective quality factor of the sample $Q= \omega T_2$, and second there is a boost by the large number of nuclei, $n_s V$ that are simultaneously being observed. We choose $^3$He because it has a fundamentally long coherence time ($T_2\approx 1000$~sec for the liquid state) and polarization of order unity has already been achieved with optical pumping techniques \cite{3He2}.

{\it Monopole-Dipole Axion Exchange.}
For concreteness, we consider a tungsten cylindrical shell of length $1$ cm, thickness $4$ mm, and outer diameter 3.8 cm divided into 20 sections of length 6 mm. The radius of each section is modulated by approximately $200$ $\mu$m in order to generate a time-varying potential at frequency $\omega = 10$ $\omega_{\rm{rot}}$, due to the difference in the axion-mediated interaction as each section passes by the sensor. The rotation of the cylinder can be accomplished by an in-vacuum piezoelectric transducer \cite{pi}. Alternatively, higher rotational speeds may be possible by using a low-friction rotary feedthru \cite{rotary} attached to an external driving motor. The rotational frequency of the cylinder can be chosen to be significantly lower than the frequency with which the axion potential is modulated. This decouples mechanical vibration from the effects of the modulated axion potential.

The $^3$He sample is contained in a quartz oblate spheroidal enclosure of internal diameters $3$ mm $\times$ $3$ mm $\times$ $150$ $\mu$m.  The minimum magnetic field that can be detected with this sample is $B_{\rm{min}}= 3 \times 10^{-19}~\frac{T}{\sqrt{Hz}}$ for $T_2=1000$ s. To allow close proximity of the source mass and detector, we assume a stretched $1$ cm $\times$ $1$ cm niobium foil screen of thickness $25$ $\mu$m covers a cutaway region of the Niobium shell between the mass and sample. The $^3$He vessel has wall thickness $50$ $\mu$m and is rigidly attached to the superconducting cylindrical shell, to minimize relative motion between the sample and shield. We assume a $50$ $\mu$m gap between the shield and rotating cylinder.

The entire NMR sample region and source mass cylinder are housed in a liquid Helium cryostat. We imagine an outer superconducting shield enclosing the apparatus screens stray background magnetic fields. We assume the rotational driving mechanism is thermally shielded and heated to operate at room temperature. If $\omega_{\rm{rot}}/2\pi = 10$ Hz and $\omega/2\pi$ is $100$ Hz, then the net $B_{\rm{ext}}$ needed at the sample is of order 30 mG. $B_{\rm{ext}}$ is the sum of the internal magnetic field of the sample, which is roughly $0.2$~Gauss for $2\times 10^{21}$ cm$^{-3}$ density of $^3$He, and a field generated by a persistent current in superconducting coils. In such a background field, we expect the SQUID can operate near its optimal sensitivity of 1.5~fT/$\sqrt{Hz}$.

\begin{figure}[!t]
\begin{center}
\includegraphics[width=1.0\columnwidth]{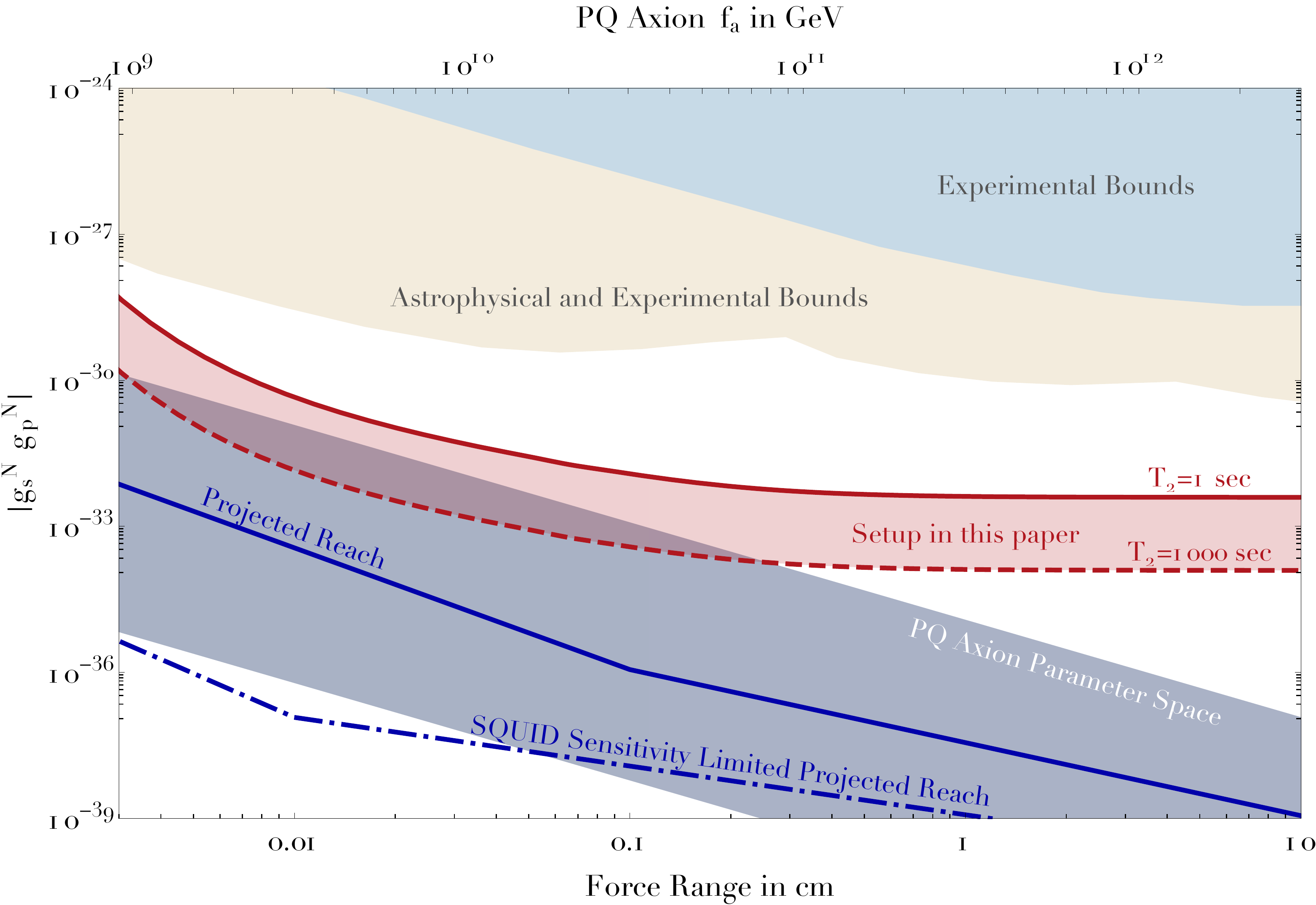}
\caption{Reach in the coupling vs interaction range plane for the monopole-dipole axion mediated interactions. The band bounded by the red (dark) solid line and dashed line denotes the limit set by transverse magnetization noise of the sample for the specific setup described in the text, for $T_2$ ranging from $1$~s to $1000$ s. The blue (darker) solid line is a future projection obtained by scaling the setup using parameters chosen in Table 1. The blue (darker) dot-dashed line is the projected limit set by the SQUID sensitivity. We limit the integration time in all setups to $10^6$~sec. The shaded band is the parameter space for the PQ axion with $C_f=1$. Additional uncertainties \cite{scalaraxioncoupling} and model dependence \cite{PDG} can produce variations of this axion parameter space. Experimental as well as combined experimental and astrophysical bounds are also presented \cite{Raffelt:2012sp,gsgpgermany}.
\label{reach}}
\end{center}
\end{figure}

In Fig. \ref{reach} we present the reach of the setup assuming a total integration time of $10^6$~sec for a monopole-dipole axion mediated interaction for both $T_2=1$ s and $1000$ s. The limitation is due to noise indicated in Eq. (\ref{bmin}), which lies significantly above the SQUID sensitivity. We also include a future projection of ultimate limits by scaling the size of the apparatus and increasing the sample density to that of liquid $^3$He. In this case, we divide our projection into two regions depending on the axion interaction range.
For axions with Compton wavelength smaller than 1~mm we assume that the sample area cannot be larger than $10^4 \times \lambda_a^2$, while it is fixed to $100$~cm$^2$ for lighter axions.
The experimental parameters are summarized in Table \ref{table1}. Finally, we draw curves for the PQ axion parameter space assuming $C_f=1$ as well as the current astrophysical and experimental bounds or a combination thereof \cite{Raffelt:2012sp}. Not only does the proposed setup compete with astrophysical bounds, but it probes a large part of the axion parameter space in the traditional axion window of $f_a$ between $10^9$ and $10^{12}$~GeV, which corresponds to axion wavelengths between $\sim 30 \mu$m and 3~cm.

\begin{table}[!t]
\begin{center}

   \begin{tabular}{@{}ccc@{}}
  \hline
  \hline
  Source Mass & Unpolarized  & Polarized\\
  \hline
W Nucleon Density& $184\times(6.4 \times 10^{22})~\mbox{cm}^{-3}$ &  \\
Xe Spin Density &&$2\times 10^{22}\mbox{cm}^{-3}$ \\
Fe Spin Density && $8 \times10^{22} \mbox{cm}^{-3} $ \\
  \hline
  \hline
  \end{tabular}
    \vskip0.2in
  \begin{tabular}{@{}ccc@{}}
  \hline
  \hline
   NMR Sample & Setup in this paper & Projected \\
  \hline
V $(\lambda_a> 1\mbox{mm})$& $(3\mbox{mm})^2 \times 150\mu \mbox{m}$ & $10^2\mbox{cm}^2 \times \lambda_a$   \\
$(\lambda_a< 1\mbox{mm})$ & same &  $10^4\lambda_a^2 \times \lambda_a$ \\
$T_2$ & $1-1000~\mbox{sec}$ &$1000~\mbox{sec}$ \\
$p$ &  1 & 1 \\
$n_s$ & $2\times10^{21}\mbox{cm}^{-3}$ & $2\times 10^{22}\mbox{cm}^{-3}$\\
\hline
\hline
SQUID $(\lambda_a< 0.1\mbox{mm})$& $1.5\frac{fT}{\sqrt{Hz}}$ &  $0.15\frac{fT}{\sqrt{Hz}} \left(\frac{1~cm^2}{10^4\lambda_a^2}\right)$\\
 $(\lambda_a>0.1\mbox{mm})$& same &  $0.15~\frac{fT}{\sqrt{Hz}}$  \\
  \hline
  \hline
  \end{tabular}

\caption{\label{table1} Summary of experimental parameters for the source mass and the NMR sample in the experimental setup described in the text and the ultimate projected sensitivity of the setup used in our estimates shown in Figs. \ref{reach} and \ref{reach-spin}.}
\end{center}
\end{table}

In order for the full sample to remain on resonance, gradients across the sample need to be controlled at the level of $\sim 10^{-11} \left( \frac{1000~\mbox{sec}} {T_2}\right)$~T. In this case, the rotation rate of the driven cylinder must also be controlled at 1 mHz to take advantage of the full $Q$. This level of control has been demonstrated for PZT driven actuators \cite{stanford}. The spheroidal shape of the sample suppresses magnetic gradients due to the magnetized gas itself. However, gradients can be produced due to image currents arising from the Meissner effect in the superconducting shield.  To minimize the effects of gradients, we assume that the $^3$He vessel has an extended length of 1 cm along the polarized (z-) direction, while the active region of the sample remains 3 mm in size. Finite element simulations indicate that the gradient is controlled in this central region at the level of $5 \times 10^{-8}$ T, which left unchecked will limit $T_2$ to approximately $1$ s for the gas density we consider.   A specially engineered superconducting coil setup can also partially cancel the gradient, allowing extension of $T_2$ up to $100$ s for a $99\%$ compensation.  However, diffusion of the gas from the central active region to and from the surrounding gradient compensation region can also contribute to decoherence. We can estimate the decoherence due to diffusion as $\exp{[-D(\gamma \nabla_z B)^2 \frac{t^3}{3}]}$, where $D$ is the diffusion constant \cite{diffusion}. Taking $D=1.7 \times 10^{-3} {\rm{cm}}^2$/s, to diffuse by 3 mm takes approximately 100 s. Thus to avoid significant mixing between the active sample region and surrounding region with larger gradients,  with a $99\%$ gradient compensation, the effective $T_2$ is reduced to approximately 10 s for a sample of the size we consider. In principle spin-echo techniques could also be employed to further reduce the effects of gradients, as in Ref. \cite{Xe-spin-echo}.

Vibrations in the apparatus e.g. due to the rotation mechanism can be a source of magnetic field noise, primarily due to the Meissner currents in the superconducting shields.  If the relative separation between the sample and the outer superconducting shield $\delta x$ varies due to acoustic vibration, the local magnetic field varies by $\frac{\partial B}{\partial x} \delta x$, where $\frac{\partial B}{\partial x}$ is the magnetic field gradient due to the presence of the image magnetization. As the cylinder rotates, we assume some wobble is possible which occurs primarily at the rotational frequency and lowest order harmonics. Although the rotational frequency of the cylinder is taken to be several times (e.g. $10$ $\times$) smaller than the magnetic resonance frequency, some vibration can in principle be transmitted at this frequency due e.g. to nonlinearities.  
Assuming a wobble in the cylinder at $\omega_{\rm{rot}}/2\pi = 10$ Hz on the order of $10$ $\mu$m, and $1$ percent of this at 100 Hz, we estimate approximately 2 nm of relative motion between the sample and outer shield. This results in a resonant magnetic field of order $10^{-22}$ T.   In addition, although we take the quartz vessel containing the sample to be rigidly attached to the inner superconducting shield, acoustic vibrations can modulate the distance between the sample and inner shield and result in magnetic noise. Assuming an amplitude of shield vibration of 2 nm, we expect relative motion between the sample and shield to which it is attached of order $10^{-17}$ m.  With a gradient of $10^{-5}$ T/m, this corresponds to a field background of $\sim 10^{-22}$ T. 

There is also a background generated by trapped fluxes in the superconducting shield. When the shield is cooled at low fields ($<10^{-10}$~T) we estimate trapped fluxes to be less than $10$ cm$^{-2}$. The thermal noise from a trapped flux in the vicinity of the sample we estimate \cite{blas-pinningforce, simanek-vortexmass} to be $7 \times10^{-20} \frac{T}{\sqrt{Hz}} \left(\frac{200~\mu m}{r}\right)^3$ where $r$ is the distance from the flux. Note that the exact properties of fluxes also depend on the shield construction and the above estimate is indicative. A $2$ nm relative motion between the sample and the outer shield introduces a coherent background field of $\lesssim10^{-22}$~T for a $10$ cm$^{-2}$ flux density.



Even if the source mass is unpolarized there is a background magnetic field oscillating at the resonant frequency due to the Barnett effect \cite{barnett}. 
The differential field from the Barnett effect as the segments of varying thickness pass by the detector will be below the $10^{-14}$ T level. This background is eliminated once the shield is placed between the source mass and the NMR sample with a screening factor $> 10^5$. This should be possible even for thin shielding layers of order tens of microns, by appropriately choosing the length of the cylindrically shaped shield that surrounds the sample and diverting stray magnetic fields around the region enclosing the sample. The sign of the Barnett effect can also be changed by reversing the rotation. The shield also attenuates magnetic noise due to thermal currents in the tungsten mass \cite{varpula}, which we estimate at $10^{-12}$ T$/\sqrt{\rm{Hz}}$.

{\it{Dipole-Dipole axion exchange}}.
In the case of a spin-polarized source mass, the source mass itself will generate (at least 0.1~Gauss) background magnetic fields fluctuating at the frequency of interest. To minimize this field, the thickness of the polarized region of the driving mass can be limited to roughly $\lambda_a$. Improved shielding factors can be achieved when the source mass is now introduced in the area of interest after the shield has gone through the superconducting phase transition. In combination with the Meissner effect, there is still the danger of locally producing fields ($> 1~T$ in the case of a ferromagnet) above the critical field. These fields can be attenuated to safe levels using a thin ($\lesssim50~\mu m$) $\mu$-metal shield layer.
In addition,
because of the strong interaction between the superconducting shield and the polarized source mass, there will be a force that will distort the shield. In the case of 200 $\mu$m thick iron magnets, this displacement can be as large as $100$ nm for a 100 $\mu$m thick shield.
For this reason, we require one more superconducting shield to be placed between the source mass and the NMR sample. Shielding will ultimately limit how close the source mass and the sample can be placed and requires special consideration when probing sub-$200$ $\mu$m distances. The ultimate reach of such a setup is estimated in Fig. 3.


\begin{figure}[!t]
\begin{center}
\includegraphics[width=0.9\columnwidth]{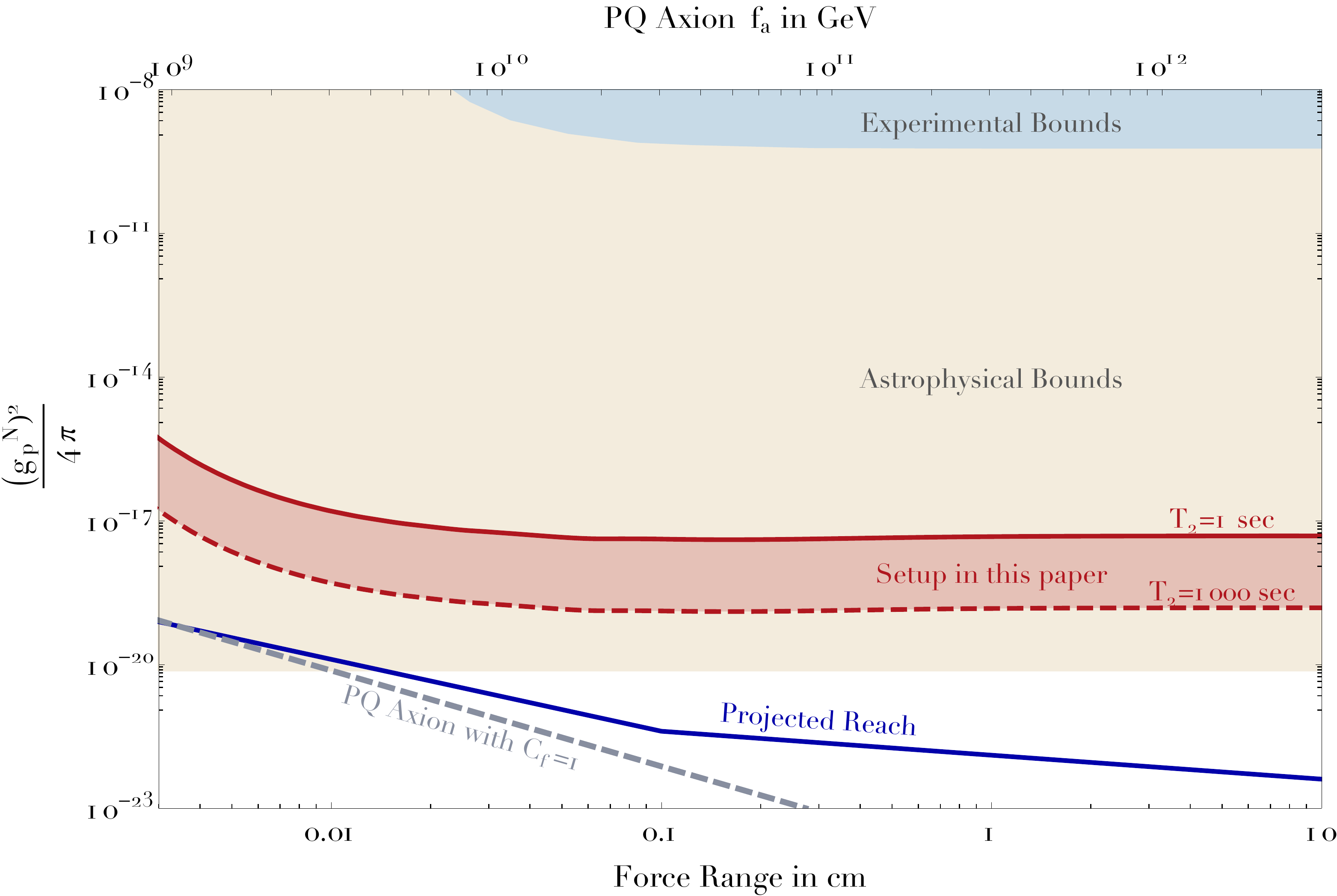}
\includegraphics[width=0.9\columnwidth]{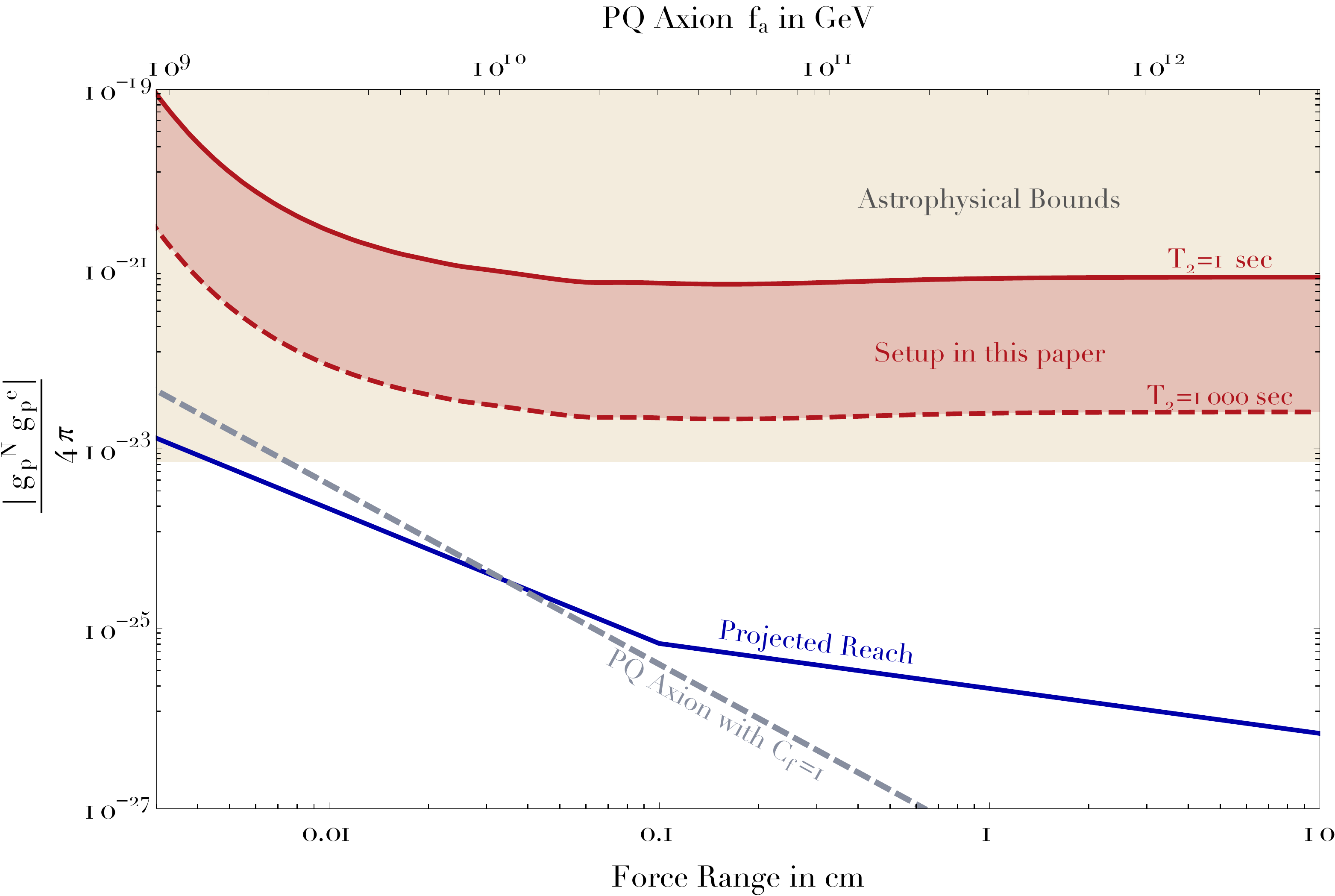}
\caption{Reach in the coupling vs interaction range plane for the dipole-dipole mediated interactions between $^3$He nuclei (top) or electrons and $^3$He nuclei (bottom). The red (dark) solid and dashed lines denote the limits set by the source mass described in the text or a liquid $^3$He sample with $T_2=1\mbox{ and }1000$~sec, respectively.  The ultimate projected sensitivity is shown with the blue (darker) solid line. Integration time is set to $10^6$~sec. Also shown is the PQ axion signal for $C_f=1$. The value of $C_f$ in specific models is discussed in Ref. \cite{PDG}. Astrophysical and experimental bounds are taken from Ref. \cite{Raffelt:2012sp}.
\label{reach-spin}}
\end{center}
\end{figure}

{\it{Discussion}}.  
When looking for dipole-dipole interactions, screening of magnetic fields in excess of 10~Gauss will be particularly challenging at sub-mm distances. These limitations mainly arise due to the mechanical stability requirements of the shield and not the screening factors needed. Sufficient screening can be achieved by a long superconducting cylinder surrounding the source mass that diverts the magnetic flux away from the NMR sample; magnetic field screening is not limited by the thickness of the shield \cite{blas-thesis}. This is only true for AC measurements such as the one described here-- for DC measurements, the amount of trapped flux has to be controlled to perhaps forbiddingly small levels.

Additional control of systematics is also possible with the use of more than one detector placed symmetrically around the superconducting cylinder screening the source mass, and testing for correlations in their signals. Specially shaped superconducting shields surrounding the sample could help reduce gradients due to the Meissner effect. While at the proposed level of sensitivity we do not expect to be limited by He collisions with the walls of the glass vessel \cite{3He2}, the use of a $^3$He and $^4$He mixture could further reduce relaxation effects induced by wall collisions in the narrow sample container \cite{mixture}.

    The method described has the potential to exceed current laboratory bounds on spin-dependent short range forces by several orders of magnitude. The $^3$He magnetic moment is dominated by the neutron contribution so it may be possible to use different NMR material to test the different combinations of nucleon interactions. Besides axions, this technique can also be used for light massive gauge boson searches -- the exact parameter space probed in this case can be extracted from our estimates of $B_{\rm{eff}}$~sensitivity. For the case of gauge bosons that kinetically mix with the photon, we estimate an ultimate reach of $\sim10^{-12}$ in the mixing at the mm range.  Most importantly,
this technique is to our knowledge the best approach so far to probing the PQ axion parameter space in the traditional axion window of $f_a$ between $10^9$ and $10^{12}$~GeV, bridging the gap between astrophysical bounds and cosmic axion searches. Eventually even the more model independent dipole-dipole mediated interaction for the PQ axion may be accessible.

{\it{Acknowledgements}}. We are grateful to Savas Dimopoulos, Blas Cabrera, and Mike Romalis for useful discussions and feedback. We also thank Aharon Kapitulnik, Eli Levinson-Falk and Benjamin Lev for their insight. We thank Jordan Stutz for finite element simulations. AA is thankful to David Cory, Peter Graham, John March-Russell, Maxim Pospelov and Giovanni Villadoro. AG is supported by grant NSF PHY-1205994. Research at Perimeter Institute is supported by the Government of Canada through Industry Canada and by the Province of Ontario through the Ministry of Economic Development and Innovation.


\end{document}